\documentclass[fleqn,usenatbib]{mnras}

\usepackage{newtxtext,newtxmath}
\usepackage[T1]{fontenc}

\usepackage{graphicx}
\usepackage{xcolor}

\title[SNe,  GRB and cocoon interactions]{The large landscape of supernova,  GRB and cocoon interactions}

\author[F. De Colle et al.]{
Fabio De Colle,$^{1}$\thanks{E-mail: fabio@nucleares.unam.mx (FDC)}
Pawan Kumar$^{2}$
and Peter Hoeflich$^{3}$
\\
$^{1}$Instituto de Ciencias Nucleares, Universidad Nacional Aut{\'o}noma de M{\'e}xico, A. P. 70-543 04510 D. F. Mexico\\
$^{2}$Department of Astronomy, University of Texas at Austin, Austin, TX 78712, USA\\
$^{3}$Department of Physics, Florida State University, Tallahassee, FL 32306, USA
}

\date{Accepted XXX. Received YYY; in original form ZZZ}

\pubyear{2021}

\begin{document}
\label{firstpage}
\pagerange{\pageref{firstpage}--\pageref{lastpage}}
\maketitle

%
%

\begin{abstract}
Long gamma ray bursts (LGRBs) are associated to the collapse of a massive star and the formation of a relativistic jet. As the jet propagates through the star, it forms an extended, hot cocoon. The dynamical evolution of the jet/cocoon system and its interaction with the environment has been studied extensively both analytically and numerically. On the other hand, the role played by the supernova (SN) explosion associated with LGRBs in determining the outcome of the system has been barely considered. In this paper, we discuss the large landscape of outcomes resulting from the interaction of the SN, jet and cocoon. We show that the outcome depends mainly on three timescales: the times for the cocoon and supernova shock wave to break through the surface of the progenitor star, and the time needed for the cocoon to engulf completely the progenitor star. The delay between the launch of the SN shock moving through the progenitor star and the jet can be related to these three timescales. Depending on the ordering of these time scales, the jet-cocoon might propagate inside the SN ejecta or the other way around, and the outcome for the properties of the explosion would be different. We discuss the imprint of the complex interaction between the jet-cocoon and the supernova shock on the emergent thermal and non-thermal radiation.
\end{abstract}


\maketitle

%
%

\section{Introduction}

Long gamma-ray bursts (LGRBs) are produced by relativistic jets ejected during the collapse of massive progenitor stars \citep[see, e.g.,][for a review]{kumar15}.
The association between LGRBs and broad-line (bl), type Ic supernovae (SNe) has been confirmed by photometric observations, with the optical light curve of LGRBs showing an increase in the flux $10-15$ days after the $\gamma$-rays trigger, and by spectroscopic observations \citep[e.g.,][and references therein]{cano17}. In addition, LGRBs are hosted by star forming galaxies \citep[][]{fruchter06} characterized by enhanced young and low-metallicity stellar populations \citep{savaglio09}.

Spherically symmetric models of Ic-bl SNe predict that they typically are $\gtrsim 10$ times more energetic than normal Ic SNe. Their absorption lines indicate fast moving material, corresponding to velocities $\gtrsim 10000-30000$ km s$^{-1}$ at the light curve peak \citep{modjaz16, ashall20}, and much larger at early times (e.g., $\sim 70000$ km s$^{-1}$ in SN 2017iuk and SN 2020bvc - see \citealt{izzo19, ho20, izzo20}). While most low redshift LGRBs in which a deep search has been done are associated to SNe, observations show that the opposite is true only in $\sim 10\%$ of type Ic-bl SNe. 

Filling the gap between energetic LGRBs and Ic-bl SNe observed without a companion GRB, intermediate class objects have been discovered recently. These include low-luminosity GRBs (ll-GRBs), which have a $\gamma-$ray luminosity of $\sim 10^{46}-10^{47}$ erg s$^{-1}$, i.e. 3-4 orders of magnitude lower than LGRBs \citep{campana06,soderberg06,pian06,starling11,margutti13}, and relativistic SNe, which are otherwise regular type Ic-bl SNe but showing in their radio emission the signature of material moving at relativistic speeds, i.e., $v_{\rm sh}\sim 0.7-0.8$ c (SN2009bb: \citealt{soderberg10}; SN 2012ap: \citealt{margutti14, milisavljevic15}).

It has been suggested that type Ic-bl SNe (without an associated GRB), relativistic SNe, ll-GRBs, X-ray flashes and luminous GRB are different aspects of the same phenomenon, that is, the result of a central compact object injecting energy into the system. While regular GRBs are relativistic jets lasting for long enough to break out of the progenitor star successfully, ll-GRBs, relativistic SNe and tye Ic-bl SNe could be associated to failed GRBs \citep{bromberg11a,nakar12} and to the GRB cocoon emission \citep{decolle18a}, or to relativistic jets seen off-axis \citep{irwin16, urata15}\footnote{Relativistic SNe, in particular, could be  accompanied by a successful, low-luminosity relativistic jets, as the coverage of the gamma-ray transient sky is incomplete (e.g., \citealt{soderberg10}).}. In the former case, the relativistic jets do not make their way through the progenitor star, and deposite all their kinetic energy in the stellar envelope. In the latter case, the highly beamed $\gamma$-ray emission is suppressed for off-axis observers, while the LGRB radio and X-ray emission should be detectable if the SN is close enough. Searches of radio emission from SNe on timescales of $\sim$ years after the explosion have failed to detect any associated off-axis GRB event \citep{berger03, bietenholz14,  ghirlanda15, corsi16}. Nevertheless, the presence of an off-axis jet has been recently inferred from the X-ray emission of SN 2020bvc \citep{izzo20}.

The GRB dynamics through the progenitor star has been extensively studied  both by hydrodynamic
\citep[e.g.,][]{aloy00, ramirez-ruiz02, zhang04, lazzati05, morsony07, wang08, mizuta09, bromberg11b, nagakura11, lazzati12, lopezcamara13, mizuta13, lopez-camara14, duffell15, hamidani17, decolle18a, decolle18b, harrison18, gottlieb20a, gottlieb21, suzuki21} and magnetohydrodynamic \citep[e.g.,][]{komissarov09, lyubarsky09, lyubarsky10, tchekhovskoy10, bromberg14, bromberg16, gottlieb20b} simulations, as well as by analytical studies (e.g.,  \citealt{bromberg11b, nakar17}).
Calculations show that the relativistic jet slows down to non-relativistic speeds as it interacts with the progenitor star.
Then, a fraction of its kinetic energy is dissipated into thermal energy. 
The post-shock region is formed by two components: the shocked stellar material, heated and accelerated by the forward shock, and the shocked jet material, heated and decelerated (in the lab frame) by the reverse shock. The lateral expansion of the shocked gas forms an extended cocoon. The two components are separated by a contact discontinuity, prone to instabilities which will lead to mixing (see, e.g., \citealt{nakar17}). Once the cocoon breaks out, the cocoon material emerges through a ``nozzle'' and expands sideways. The four-velocity stratification with the polar angle is determined by the acceleration out of the nozzle and the rarefaction wave that propagates through the outflowing cocoon material. The cocoon will then  quickly engulfs the progenitor star, then expanding at relativistic to sub-relativistic speeds (depending on the polar angle).

While the propagation of the jet, its interaction with the stellar envelope and the formation of the cocoon have been studied in detail, the role played by the SN in determining the final outcome of the system has not been typically considered, except in a small number of cases.  \citet{ramirezruiz10} studied numerically the interaction between the SN ejecta and the GRB-jet over timescales of $\sim$ years, assuming that they evolve previously as two independent components, and showing that the GRB ejecta is swept up by the more energetic SN ejecta.  \citet{margalit20} assumed that, once the relativistic jet slows down at large distances ($r\approx 10^{16}-10^{18}$ cm, depending on the density of the circumstellar medium), it spreads laterally engulfing the supernova. They computed the radio emission resulting from the interaction between the SN-remnant and the cocoon associated with the relativistic GRB-jet, and suggested that the radio emission from the SN is strongly suppressed while the SN remains inside the cocoon. According to them a radio flare is produced several years after the explosion when the SN breaks out from the cocoon.

While these previous works studied the late-time interaction of the GRB-jet with the SN remnant, \citet{decolle18a} recognized that, if the relativistic jet breaks out first from the star, the SN shock front will then propagate through the GRB-jet cocoon instead of  through the wind of the progenitor star and that affects the dynamics and the radiation we see.

In this paper we show that it is indeed the early interaction (inside the progenitor star and just after breakout) between the SN-outflow, the GRB-jet cocoon, the progenitor star and the environment which is crucial to determine the outcome of the system and its large scale evolution.
We show that this interaction leads to different outcomes, depending on the SN and GRB energies, and on the time lag between the ejection of the GRB-jet and the SN. Guided by a set of numerical simulations, we give here a general picture of the system. More detailed calculations are left to a future study.

This paper is structured as follows: in Section 2 we present the general description of the system. In Section 3 we discuss the implications for the emission properties of GRBs and the SNe powered by a central engine. Finally, in Section 4 we drive our conclusions.

\section{General description of the system}

In this section, we present a general, qualitative description of the rich landscape resulting from the interaction between the relativistic jet, the associated SN and the progenitor star. First, we consider the SN and GRB dynamics separately. Then, we discuss the outcome of their interaction. A series of numerical simulations of the propagation of the relativistic jet and the SN through the progenitor star and its wind will guide our discussion. 

The two-dimensional, axisymmetric simulations\footnote{Three dimensional numerical simulations change quantitatively the dynamics of the jet (see., e.g., \citealt{harrison18}) but not the general picture outlined in this paper.} employ the special relativistic hydrodynamics code \emph{Mezcal} \citep{decolle12a}. The initial conditions for the simulations are the following: the SN explosion is initialized by using a ``piston'' model, i.e. by injecting a certain amount of energy $E_{\rm SN}$  and mass $M_{\rm SN}$ ($=0.1$ M$_\odot$) from a spherical region with radius $r_{\rm SN} = 5\times 10^8$ cm. This energy is injected during 1 s by assuming a velocity $v=\sqrt{2 E_{\rm SN}/M_{\rm SN}}$. 
From the same radius, we inject two relativistic jets (but only the one propagating with positive velocities is simulated), with a total energy E$_j=2\times 10^{51}$ erg, a Lorentz factor $\Gamma_j = 10$, an opening angle $\theta_j = 0.2$ rad, and with a duration $t_j=20$ s, so that the jet luminosity is given as $L_{\rm j}=E_{\rm j}/t_{\rm j} = 10^{50}$ erg s$^{-1}$. These values are typical of LGRBs.
The jet and the SN are launched into a massive, compact Wolf-Rayet progenitor. We employ the E25 model of \citet{heger00} This is a 25 M$_\odot$ star which has lost most of its mass by winds. The final mass of the progenitor is $\sim 5$ M$_\odot$ and its final radius is $\sim 3\times 10^{10}$ cm. Outside the stellar envelope, we set the ambient density by considering a typical Wolf-Rayet wind. Thus, the density is given as $\rho= \dot{M_w}/(4\pi r^2 v_w)$, with $v_w=10^8$ cm s$^{-1}$ and $\dot{M} = 10^{-5}$ M$_\odot$ yr$^{-1}$. In the initial conditions, we neglect the wind velocity, as it is much smaller than the jet and cocoon velocities.
We use a computational box with size $L_r\times L_z$. 

To illustrate the different outcome resulting from the interaction between the SN and the relativistic jet, we run two simulations with a small computational box size ($L_r=L_z = 1.2\times 10^{11}$ cm), in which the supernova (spherical) has an energy $E_{\rm SN}=10^{52}$ erg and $E_{\rm SN}=4\times 10^{52}$ erg respectively. In this simulations, we employ a grid with 40$\times$40 cells, resolved with 9 levels of refinement. Then, the minimum cell size is $1.2\times 10^7$ cm. To describe the early phases of interaction between the SN and the cocoon, we run a simulation with a larger computational box ($L_r=L_z = 3\times 10^{12}$ cm). In this case, we consider a lower supernova energy $E_{\rm SN}=4\times 10^{52}$~erg, injected asymmetrically, i.e. with a density of the material injected from the inner boundary scaling as $\cos(\theta)^2$, being $\theta$ the polar angle.
Our coarsest grid has 40$\times$40 cells and we use 15 levels of refinement, corresponding to a maximum resolution of $4.5\times 10^6$~cm.

\begin{figure*}
\centering
 \includegraphics[width=1\textwidth]{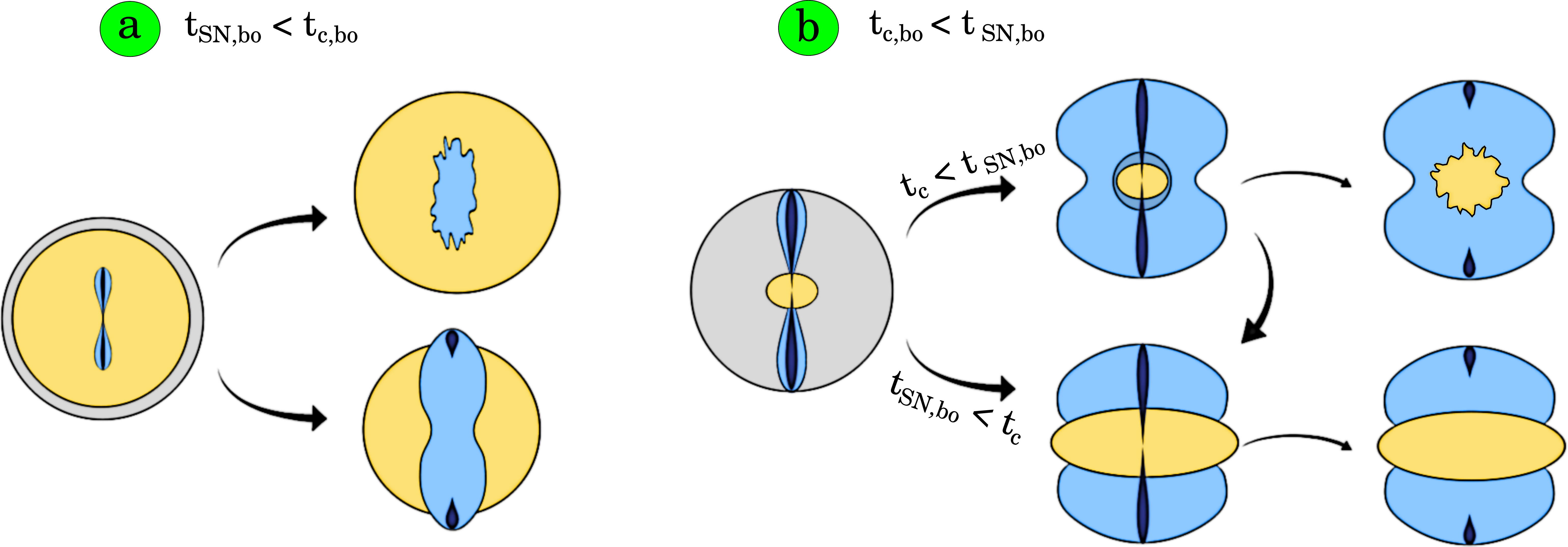}
 \caption{Schematic representation of the outcome resulting from the interaction of the supernova (yellow), the relativistic jet (dark blue), the jet cocoon (light blue), and the progenitor star (gray). The outcome depends mainly on the time needed for the cocoon and the SN to break out from the progenitor star ($t_{\rm c, bo}$ and $t_{\rm SN, bo}$ respectively), and on the time needed for the cocoon to completely engulf the progenitor star ($t_{\rm c}$). In case a), the SN breaks out first. Then, the GRB might break out from the SN at larger distances, or fail to do so. In case b), the cocoon breaks out first, associated with a successful or a choked jet. If the SN shock front breaks out before the cocoon engulfs completely the star, it expands mainly along the equatorial direction while the cocoon occupies the polar direction. Otherwise, the SN shock front breaks out into the GRB cocoon, expands and might or not break out from it at late times.}
 \label{fig1}
\end{figure*}

\subsection{Relativistic jet propagation}

First, we briefly discuss the GRB propagation through the environment. In the collapsar model \citep{woosley93,macfadyen99}, the relativistic jet is produced in a collapsing core which angular momentum is large enough to form a disk and a black hole (BH). Then, the relativistic jet is driven by magnetic forces, the rapid BH rotation and/or neutrino annihilation on the symmetry axis. Alternative models invoke the presence of magnetars. 

As the jets drills through the envelope of the progenitor star, it slows down. In the hydrodynamics case, the shock velocity can be estimated by considering the balance between the ambient and the jet ram pressures, as seen in the shock frame, that is \citep[e.g.,][]{begelman89, matzner03, bromberg11b, decolle12a}
\begin{equation}
      \rho_j h_j \Gamma_j^2 \Gamma_h^2 (\beta_j-\beta_h)^2 + P_j = \rho_a h_a \Gamma_h^2 \beta_h^2 + P_a \;,
  \end{equation}
  where the suffixes $j,a,h$ refer to the jet, the ambient medium and the jet head respectively, $h=1+P/(\rho c^2)$ is the specific enthalpy, and $\rho$, $P$, $\Gamma$, $\beta$ are the density, pressure, Lorentz factor and velocity.
  By assuming that the shock is strong (i.e., $P_a\ll \rho_a\Gamma_h^2$), we get  
  \begin{equation}
      \beta_{\rm sh} = \frac{\beta_{\rm j}}{1+(\rho_a/\Gamma_j^2 \rho_j h_j)^{1/2}} \;.
      \label{eq:vel}
  \end{equation} 
In the dense, inner core of the star, $\eta = \rho_a/ (\Gamma_j^2 \rho_j h_j) \gtrsim 1$, and the jet moves at sub-relativistic speeds, i.e. $\beta_{\rm sh} \sim \eta^{-1/2}$ (above the recollimation shock, $\Gamma_j\sim 1/\theta_j$, see \citealt{bromberg11b, harrison18}). As the jet expands through the star, the head velocity increases, achieving eventually mildly relativistic speeds in extended, low-mass envelopes, while remaining newtonian in compact progenitors (see \citealt{bromberg11b} for a detailed discussion).

Given the breakout time \citep{harrison18}
\begin{equation}
  t_{\rm j,bo} \sim 17 L_{\rm jet,50}^{-1/3} \theta_{10^\circ}^{4/3} R_{11}^{2/3} M_{10}^{1/3} \;  {\rm s},
\end{equation}  
the jet breaks out of the star if lasts long enough, i.e. $t_j\gtrsim t_{\rm j,bo}-R_\star/c$, being $R_\star$ the stellar radius. Otherwise, the reverse shock crosses the jet material at a velocity $\sim c$, and the jet is chocked inside the star. 
Then, most of the jet kinetic energy, $E_{c,i} \sim L_j t_{\rm j,bo}\sim 10^{50}-10^{51}$ erg s$^{-1}$ or $E_{c,i} \sim L_j t_j$ for successful and failed jets respectively, is dissipated into thermal energy, producing a hot, extended cocoon component which expands through the progenitor star mainly along the polar direction. 

As the jet breaks-out from the stellar surface, the cocoon quickly expands around the progenitor star at a velocity $v$ ($\sim$ a fraction of the speed of light). Then, in a time 
\begin{equation}
  t_{\rm c} \sim \frac{\pi R_\star}{2v} +  t_{\rm j,bo}\sim 10-20 \frac{R}{R_\odot} {\rm s} + t_{\rm j,bo}
\end{equation}
the cocoon shock front arrives to the equatorial plane engulfing completely the star.

The outer cocoon velocity, energy and densities are strongly stratified both along the radial and polar direction. The shock velocity is highly relativistic (with a Lorentz factor of $\sim$ 100) close to the jet core, and drops to sub-relativistic speeds in the equatorial plane (e.g.,  \citealt{lopezcamara13, gottlieb21, suzuki21}). The same steep gradient is present in the energy and density distribution.

An order of magnitude estimation of the mass and energy of the cocoon can be obtained by the following argument (see, e.g., \citealt{nakar17}).
The cocoon mass is approximately given as the mass shocked by the jet while moving inside the progenitor, i.e. by a fraction $\Omega_j/4\pi$ of the stellar mass, where $\Omega_j\simeq 2\pi \theta_j^2$ is the jet solid angle. Then,  the mass of the cocoon is 
\begin{equation}
 M_{c,i} \gtrsim 0.05 M_\odot \; \theta_{j,0.1}^2 M_{10}  \;,
\end{equation}
for a star of 10 $M_\odot$ and a jet with an opening angle $\theta_j=0.1$ rad. On the other hand, the cocoon energy is given by the energy deposited by the jet while crossing the star at  non-relativistic speeds. That is, 
\begin{equation}
    E_c \sim   t_{\rm j,bo}  L_{\rm jet} \sim 1.7\times 10^{51} L_{\rm jet,50}^{2/3} \theta_{10^\circ}^{4/3} R_{11}^{2/3} M_{10 }^{1/3} \; {\rm erg \; s}^{-1}\;.
\end{equation}

\subsection{Supernova propagation}
\label{sec:SN}

The GRB cocoon energy is typically large enough to unbind the progenitor. Nevertheless, the GRB itself can not produce amounts of $^{56}$Ni large enough to explain the observations (i.e., $\lesssim 0.5$ M$_\odot$). In fact, as mentioned above, the amount of stellar mass shocked by the jet solid angle $\Omega_j/4\pi \sim 5\times 10^{-3} (\theta_j/0.1)^2$ corresponds to a small fraction of the stellar material\footnote{In a recent paper, \citet{barnes18} obtained 0.24 M$_\odot$ of Ni$^{56}$ by the interaction of a relativistic jet with the stellar material, in apparent contradiction with the simple analytical argument presented here. Actually, their figure 2 seems to indicate that the Ni$^{56}$ is somehow produced in the jet material, which mass available is much smaller than the one needed to power a SN.}, implying that  $M_{^{56} \rm Ni}\lesssim M_\star \Omega/4\pi \approx 0.05$ (M$_\star$/10 M$_\odot$) M$_\odot$. On the other hand, the cocoon shock expanding inside the progenitor star is not strong enough to produce a large amount of $^{56}$Ni, and leads mainly to the production of intermediate mass elements (e.g. $^{28}$Si, $^{32}$Ca, etc, see, e.g., \citealt{tominaga07, maeda09}).

Among other possibilities, the SN shock front could originate into a wind ejected from the collapsar disk \citep{macfadyen99}, or the SN can be associated to energy dissipation driven by a strongly magnetized, rapidly rotating magnetar \citep{bucciantini07,bucciantini09, metzger11, metzger15}, or it could be associated to the jittering jet mechanism (see, e.g., \citealp{papish11, papish14}).
Typically, for core collapse SNe, about $\approx 0.07-0.3 M_\odot$ of $^{56}$Ni are produced \citep{anderson19,davis21}. From the nucleosynthesis and, in particular, for large $^{56}$Ni masses, the explosion  must be aspherical \citep{thielemann19} which may be hard to explain within the framework of 'classical' neutrino driven explosions in non-rotating cores which seem to work well in lower mass progenitors \citep{orlando21}.

The SN shock front will break out from the stellar surface in a timescale $t_{\rm SN,bo} \sim$ tens of seconds for compact Wolf-Rayet progenitors leading to type Ic SNe, where the exact value depends on the SN  energy, and on the stellar structure and radius. The density of the expanding SN, when breaking out of the star, is strongly stratified, with $\rho\propto r^{-n}$ and $n\approx7-10$ \citep[e.g.,][]{tan01}.
Thus, while most of the mass (and energy) in the SN is moving at $\sim 10^4-2\times 10^4$ km s$^{-1}$, a fraction of the energy ($\sim 10^{46}-10^{47}$ erg) is moving at larger speeds ($\sim 0.1-0.2$~c). In typical SNe type Ib/c, the SN shock front moves through the  
wind of the progenitor Wolf-Rayet star, which typically has 
$\dot{M}\sim 10^{-5}-10^{-6}$ M$_\odot$ yr$^{-1}$, and its velocity drops with 
time as $v_{\rm sh} \propto t^{m-1}$, with $m=(n-3)/(n-2)\sim 0.8-0.9$ \citep{chevalier82}.
Electrons accelerated by the fast moving shock emit synchrotron radiation, observed in radio, while the bulk of the SN ejecta is responsible for the thermal optical emission commonly observed in SNe.

\subsection{Interaction between the supernova and the relativistic jet}

Next, we discuss how this general picture is modified by the interaction between the SN shock front, the GRB cocoon and the progenitor star.
It is unclear if the SN is launched before the GRB (or viceversa). Observations are not very constraining,  e.g. $\lesssim 4$ ks delay for the SN 2006aj associated to GRB 060218 \citep{campana06}. In the collapsar scenario, if the relativistic jet is launched once the proto-neutron star collapses to a BH, a delay of a few seconds is possible, with the SN launched from the standard neutrino mechanism or from the accretion disk wind \citep[see, e.g.][]{obergaulinger17, aloy18}. The jet can be delayed by a few seconds also in the case of a magnetar \citep[e.g.,][]{bucciantini07, bucciantini09} because of neutrino driven wind is much more heavily baryon loaded for a few seconds thereby limiting the jet speed, or in the collapsar scenario, as long as the ram pressure of the collapsing material is larger than the jet ram pressure.

Figure 1 shows a schematic representation of the complex dynamics resulting from this interaction. 
As we will discuss in some detail in the following, the evolution of the system is mainly regulated by three timescales: 1) $t_{\rm c,bo}$, the cocoon break-out time from the progenitor star, which coincides with the jet break out time for successful GRBs; 2) $t_{\rm SN,bo}$, the SN break-out timescale; 3) $t_{\rm c}$, the time needed for the GRB cocoon to completely engulf the progenitor star.
As we discussed above, these parameters depend on the time delay between the onset of the SN shock front and the GRB ejection, on the amount of energy associated with the SN, on the progenitor star radius and mass, and on the characteristics of the GRB (opening angle, luminosity, presence of a dynamically important magnetic field, etc).

\begin{figure}
\centering
 \includegraphics[width=0.4\textwidth]{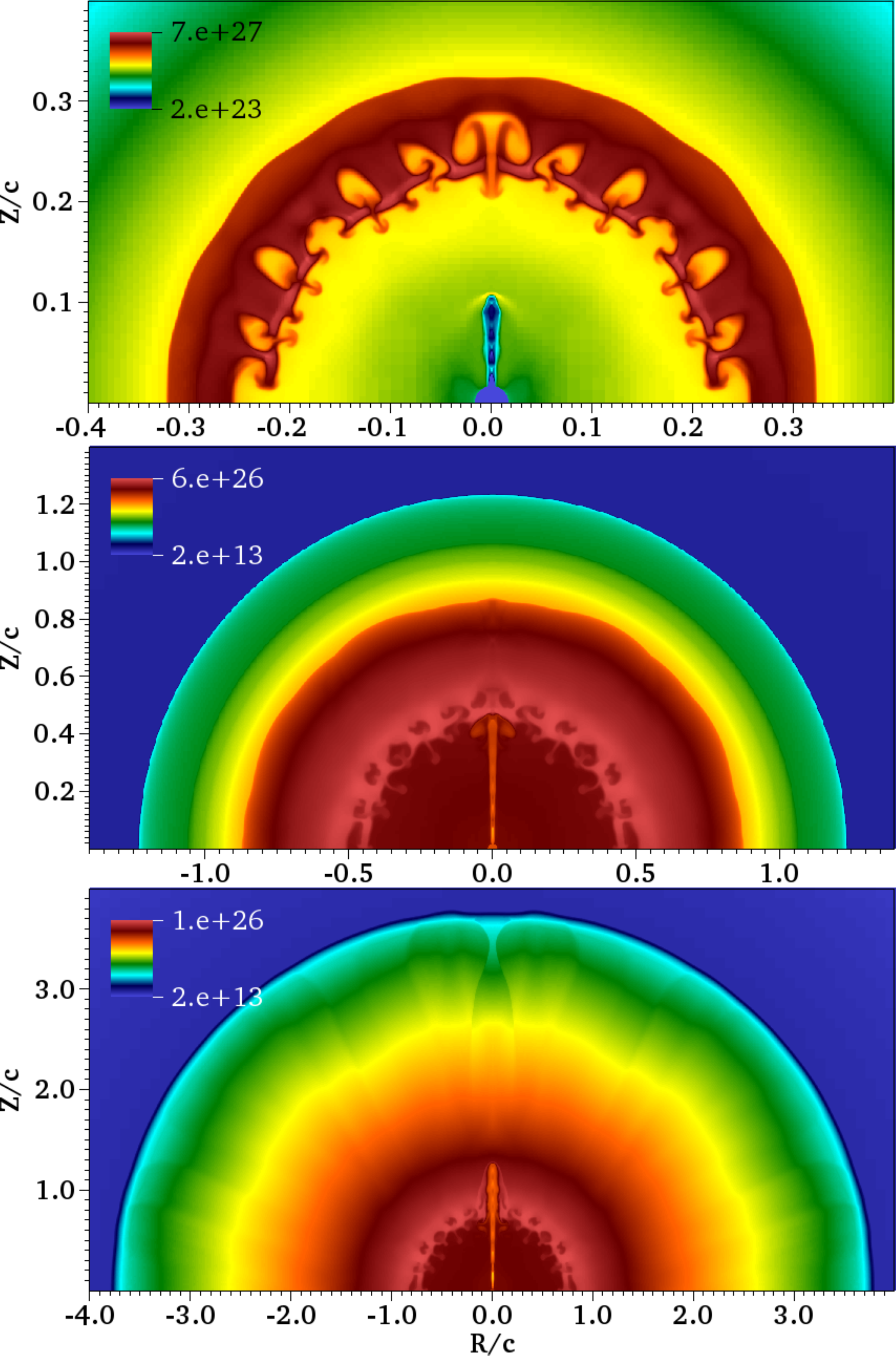}
 \caption{Number density maps at 2.25 s, 4.5 s and 7.5 s (from top to bottom) illustrating the scenario shown in the panels a) of figure \ref{fig1}. Lengths are in units of the speed of light, while the number density is in units of cm$^{-3}$. A spherical SN is injected from an inner boundary located at $5\times 10^8$ cm, with an energy $E_{\rm SN} = 4\times 10^{52}$ erg, and a mass $M_{\rm SN} = 0.1 M_\odot$.
 The grid has a size $L_r=L_z=1.2\times 10^{11}$ cm, with $40\times40$ cells at the coarsest level of refinement, and 9 levels of refinement, corresponding to a resolution of  $1.2\times 10^7$ cm.
 The jet, injected 2 s after the SN explosion from the same location as the SN, with an energy of $10^{51}$ erg, does not overcome the SN shock front while it is still inside the progenitor star. Thus, the SN shock front breaks out first, expanding into the wind of the Wolf-Rayet progenitor. The GRB, then, moves through the highly stratified SN ejecta. At late times, the relativistic jet will eventually be chocked or break out from the SN ejecta. A similar outcome is expected in SNe with lower energies if associated to GRBs with lower luminosities and/or launched with a larger delay with respect to the SN.}
 \label{fig2}
\end{figure}
\begin{figure}
\centering
 \includegraphics[width=0.4\textwidth]{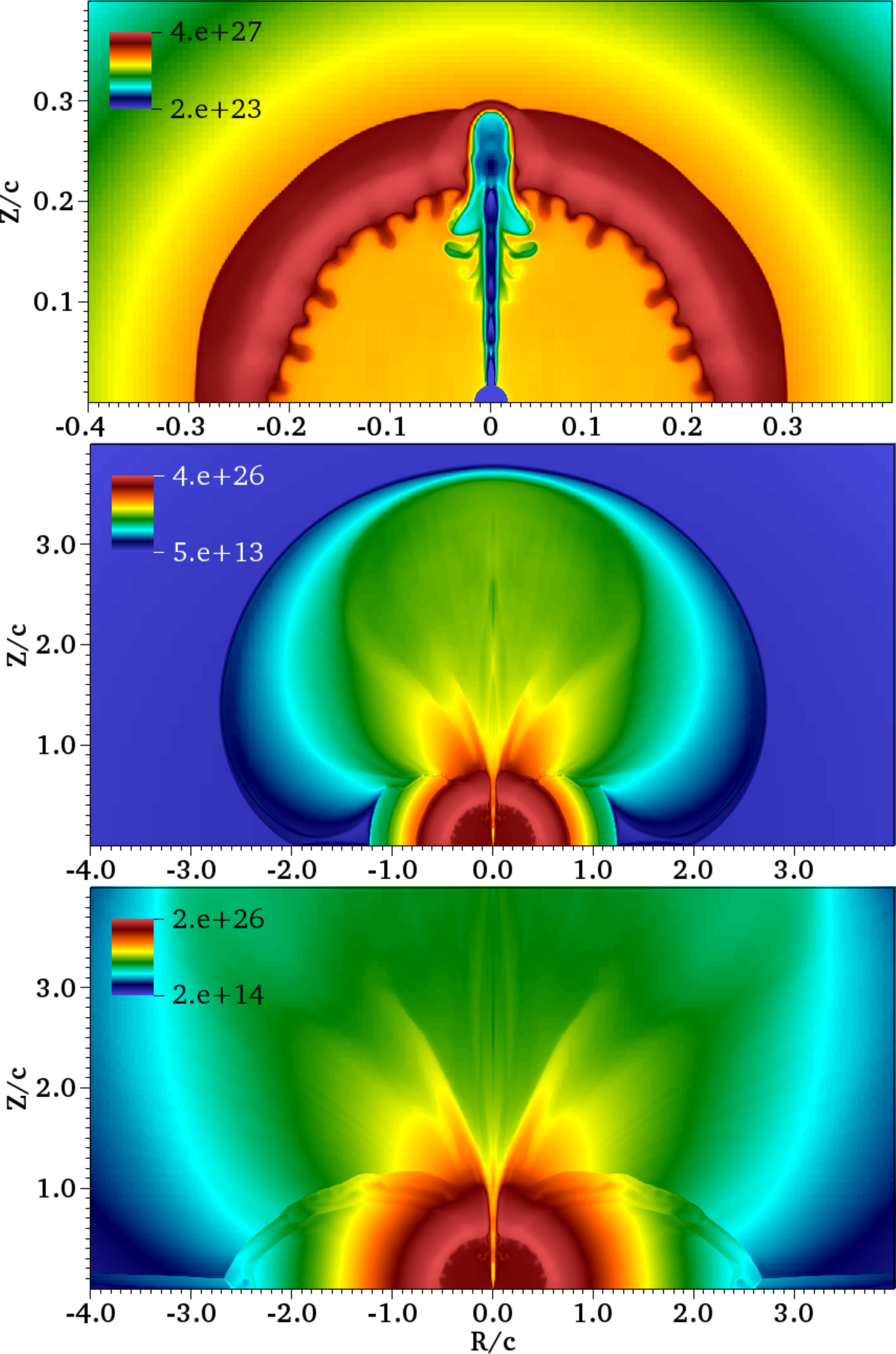}
 \caption{Number density maps at 4.5 s, 10 s, 13 s (from top to bottom). 
 Lengths are in units of the speed of light, while the number density is in units of cm$^{-3}$. 
 The numerical simulation illustrates the scenario shown in the panels b) of figure \ref{fig1}.
 The initial conditions of the simulation are the same as those shown in figure \ref{fig2}, except for the SN energy which is lower in this case ($E_{\rm SN} = 10^{52}$ erg).
  The jet overcomes the SN shock front while still inside the star. Then, it breaks out first. The jet cocoon engulfs completely the star. The SN ejecta breaks out into the cocoon material (denser than the wind of the Wolf-Rayet star) in a strongly asymmetric way, and expands mainly along the equatorial plane. The case of the SN breaking out from the star before the cocoon engulfs completely the star is also possible, and it is described in the text.}
 \label{fig3}
\end{figure}

\subsubsection{Case $t_{\rm SN,bo} < t_{\rm j,bo}$}

The left side of figure \ref{fig1} shows the case in which the SN breaks out of the star before the GRB ($t_{\rm SN, bo}<t_{\rm j, bo}$). This corresponds to the case in which the SN is launched long enough before the jet or if they are launched at a similar time but the SN luminosity per unit solid angle $\sim 8\times 10^{50} E_{52}/t_{\rm SN}$ erg s$^{-1}$ sr$^{-1}$ (being $t_{\rm SN}$ and $E_{52}$ the SN injection time in seconds and energy in units of 10$^{52}$ erg) is much larger than the corresponding GRB luminosity $ \sim 8\times 10^{50} E_{51}/ (\theta_{j,0.1}t_{\rm j,20})$ erg s$^{-1}$ sr$^{-1}$  (being $t_{\rm j,20}$ the jet duration in unit of 20 seconds and $ \theta_{\rm j,0.1}$ the jet opening angle normalized to 0.1 rad). If the SN and jet luminosity per unit solid angle are similar (or the SN luminosity is slightly larger), the jet will arrives first to the stellar edge as the collimation of the jet by the cocoon will significantly increase its propagation velocity.

In this case, the SN shock break-out and dynamics through the progenitor wind are like those described in a SN without an associated GRB (see Figure \ref{fig2}). On the other hand, the propagation of the jet is initially faster than in the progenitor star as the expansion of the SN shock wave drops the density of the environment at small radii. In the simulation shown in Figure \ref{fig2}, for instance, the density stratification in the progenitor star is $\rho_{\rm SN} \approx 10^{6} (r/5\times10^8\; {\rm cm})^{-2.75}$ g cm$^{-3}$. This is the density of the medium that the jet should have crossed if the SN was not present. In our simulations, once the SN moves through the medium, it leaves behind a cavity with a density of $\sim 1-10$ g cm$^{-3}$, i.e. a factor $\sim 10^2-10^5$ smaller than the density of the star at similar radii. As stellar material is dragged by the SN shock front, at the edge of the SN cavity the SN ejecta achieves densities much larger than the stellar density at the same radius, e.g. $3\times 10^3$ g cm$^{-3}$ vs $2\times 10^2$ g cm$^{-3}$ at $r\approx 10^{10}$ cm for the simulation shown in Figure \ref{fig2}.

As a consequence, the jet moves nearly at its injection velocity (i.e. at highly relativistic speed) inside the SN cavity, arriving to its edge in a time $t_i\sim R/c$. The position $R$ where SN and jet interact with each other depends mainly on the time delay $\Delta t$ between SN and jet injection, i.e. $R=v_{\rm SN} \Delta t/(1-v_{\rm SN}/v_{\rm j})\sim v_{\rm SN} \Delta t$. This radius is inside the stellar radius if $\Delta t\lesssim 23 (v_{\rm SN}/0.1 {\;\rm c})(R_\star/R_\odot)$ s. Theoretically, the delay between SN and GRB is expected to be smaller than this value, so that the interaction between the two shocks should always happen inside the star.

Once the jet collides with the SN ejecta, it decelerates to non relativistic speeds (e.g. to $\sim 0.05$ c in the simulations shown in Figure \ref{fig2}).
Then, in $\sim$ several seconds it will break out from the SN ejecta, as long as the SN ejecta still moves deep inside the star. Once it approaches the stellar edge, where the density stratification drops faster than $r^{-3}$, the SN shell expands radially, with the shock front accelerating up to mildly relativistic speeds. Then, the jet will moves through the highly stratified expanding material of the supernova instead of the windy material surrounding the Wolf-Rayet progenitor, and will remain confined inside the SN ejecta. 

We expect that the jet will break out from the SN in the following cases: 1) the jet head is very close to the SN shock front when the SN breaks out from the star; 2) the jet duration is very large, such that the jet has enough time to move through the SN ejecta\footnote{But, it is important to notice that although the density of the expanding SN ejecta is lower than the density of the progenitor star, its pressure increases in the region close to the jet injection. Thus, long-lasting, low luminosity jets can not be launched successfully, as the jet velocity at the injection point must be larger than the sound speed of the medium \citep[see the discussion by][]{aloy18}.}; 3) the delay between the SN explosion and the relativistic jet injection is very large (with a jet e.g. driven by material falling-back onto the central engine), such that the SN ejecta has expanded substantially, its density has dropped enough and the jet moves with relativistic speed inside the SN. In all other cases, we expect the jet to be drown inside the expanding SN. If the jet breaks out successfully, the SN shocked material will occupy most of the solid angle of the expanding ejecta, with the GRB (and its associated cocoon) moving with high relativistic speed along the polar direction.
The detailed dependence of these scenarios on the jet/SN physical parameters and the possible consequences on the observed SN light curve and spectra are left for a future study.

\subsubsection{Case $t_c < t_{\rm SN,bo}$}

Panel b) of Figure \ref{fig1} illustrates the outcome of the system when the jet reaches the SN shock front deep inside the star, and the jet and/or cocoon\footnote{Relativistic jets lasting $t_j\lesssim t_{\rm c,bo}$ will be chocked inside the star, forming in this case a so called ``failed GRB''.} break out first from the progenitor star (i.e., $t_{\rm c,bo} < t_{\rm SN,bo}$), followed by the SN shock front.

The jet breaks out first, and the cocoon expands laterally around the star, eventually engulfing it in a time $t_c$. Then, there will be two cases potentially leading to very different outcomes. If the SN shock front arrives to the stellar edge before the cocoon completely engulfs the star (i.e., $t_{\rm SN,bo} < t_c$), it breaks out and it expands through the wind of the progenitor star. As the highly pressured cocoon is expanding laterally inside the star, its pressure prevents the vertical expansion of the SN ejecta, which breaks out mainly along the equatorial direction. Then, the SN and cocoon shock fronts will expand side to side. Mixing along the discontinuity between the high entropy cocoon and the dense SN ejecta will occur. 

On the other hand, if the cocoon completely engulfs the star before the SN breaks out from it, the SN breaks out into the jet cocoon (see Figure \ref{fig3}).
Then, the SN shock front will move inside the jet cocoon. As we will show in the following, the evolution of the SN shock front is strongly dependent on the radial density and velocity stratification of the cocoon.

As discussed above, as the SN moves inside the star, it is pushed sideways by the expansion of the jet cocoon, remaining confined into a region with an $\approx$ ellipsoidal shape while moving inside the star. This, together with the cocoon density which is also strongly stratified along the polar direction, results into a highly asymmetric SN break out (see Figure \ref{fig3}, bottom panel).

\begin{figure}
\centering
 \includegraphics[width=0.37\textwidth]{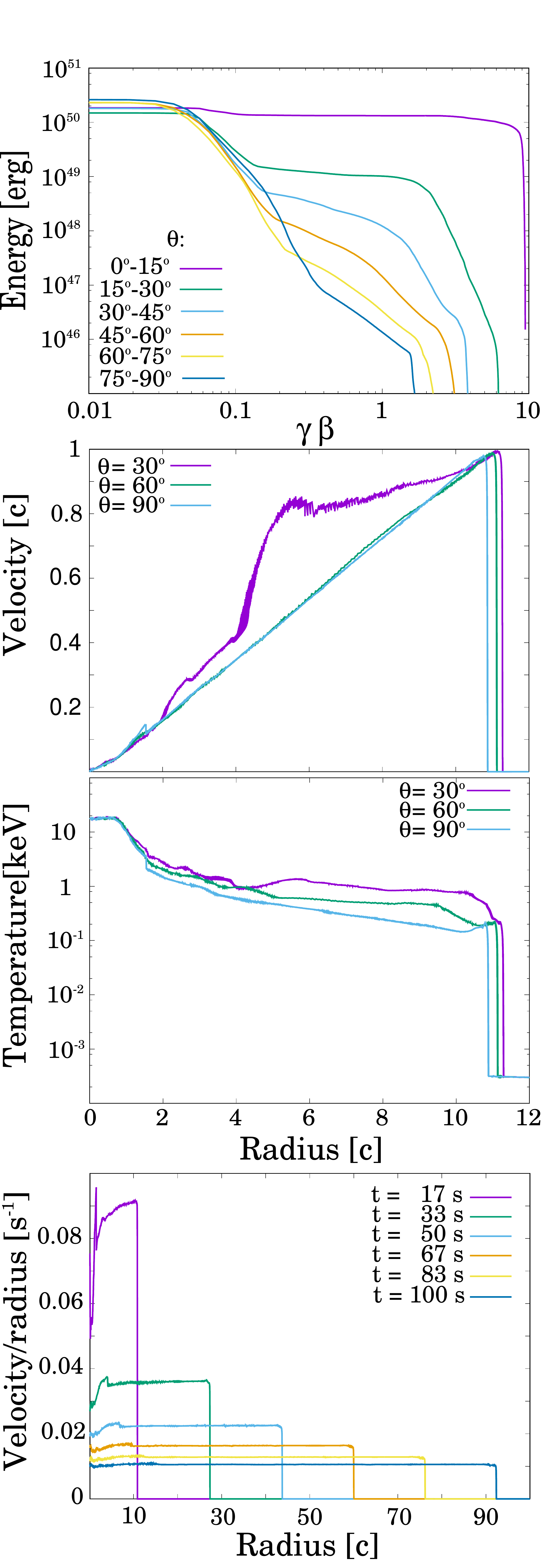}
 \caption{Energy, velocity and temperature profiles as a function of radius and time. In all the panels, the lengths and velocities are normalized to the speed of light. An aspherical SN is injected from the same inner boundary as the simulations presented in figures \ref{fig2} and \ref{fig3}, with an energy scaling as $\cos^2\theta$, being $\theta$ the polar angle, and a total energy of $4\times 10^{51}$ erg. As in the previous simulations, the jet is injected 2 s after the SN explosion. The grid has a size $L_r=L_z=3\times 10^{12}$ cm, with $40\times40$ cells at the coarsest level of refinement, and 15 levels of refinement, corresponding to a resolution of $4.6\times 10^6$ cm.
 \emph{Top panel}: Integral of the energy over velocities larger than $\gamma\beta$, as a function of $\gamma \beta$, for different polar slices (at 50 s). \emph{Central panels}: Velocity (normalized with respect to the light speed) and temperature profiles, at 17 s and shown for different polar angles. \emph{Bottom panel}: velocity over radius as a function of time, seen at $\theta=90^\circ$. }
 \label{fig4}
\end{figure}

Figure \ref{fig4} (upper panel) shows the energy as a function of the velocity 4-vector ($u = \gamma \beta$) and integrated over different solid angles. Close to the $z$-axis, most of the energy is carried by material moving at large speed, i.e. the relativistic jet. At larger polar angles we have the cocoon material which moves at sub-relativistic velocities, and the SN shock front expanding at a fraction of the speed of light ($v\sim 0.1-0.3$ c). The separation between the SN and cocoon components is clearly visible in the top panel of figure \ref{fig4} as a break in the decay of energy vs $u$ curves. 

The middle two panels of Figure \ref{fig4} show the velocity and  temperature profiles at 17 s along different polar angles.  While at $\theta=30^\circ$ the velocity profile is irregular (due to the contribution of the deepest portion of the jet cocoon), at larger polar angles it acquires a typical $v\propto r$ structure characteristic of homologous expansion. The entropy rich cocoon material has a temperature in the keV energy range (thus, we expect it will emit soft X-rays, as we will discuss below), nearly independent of the polar angle. 
The thermal energy density in the cocoon -- which is created by the GRB relativistic jet moving through the progenitor star -- is larger close to the jet axis and falls-off at larger angles. Thus, the expansion of this material after break-out leads to the temperature gradients seen between R $\sim$ 5 c and R $\sim$ 10 c, with the temperature of the cocoon at 30$^\circ$ being a factor of $\sim 5-10$ larger than the temperature at 90$^\circ$.

The cocoon, with a lower energy than the SN ($\approx 10^{50}-10^{51}$ ergs vs $\lesssim 10^{52}$ ergs), decelerates faster.
Close to the jet axis, the jet/cocoon velocity is much larger than the SN shock front velocity. In addition, the pressure difference between these two regions will make impossible for the SN ejecta to break out from the cocoon along the polar direction. Along the equatorial direction, the situation is more complex. 
As the SN shock front moves through the cocoon, it might arrive to the location in which the SN velocity is of order of the cocoon velocity. Then, turbulence at the interface between the SN shock front and cocoon material will mix the two components (see figure \ref{fig1}). In this case, the density structure of the SN and cocoon system will be different from the $\rho\propto r^{-n}$ solution predicted by self-similar solutions of expanding SNe (see Section \ref{sec:SN}). This is clearly illustrated in the second and fourth panels of figure \ref{fig4}. The SN shock front, in particular, is visible as a small velocity jump at $R\approx 1.8$ c in the second and fourth panels. During this phase, the velocity difference between the SN and cocoon is small $\lesssim 15000$ km s$^{-1}$. Subsequently, the velocity difference quickly disappears leading to mixing at the interface between the SN ejecta and the cocoon (see the bottom panel).

As the energy is injected in the computational box with a $\cos^2(\theta)$ profile in the simulation shown in figure 4 (with respect to a spherical injection in the simulation shown in figures \ref{fig2} and \ref{fig3}), the SN energy is lower along the equatorial plane in the former case. 
Thus, we suggest that a different outcome, with a much stronger SN shock propagating through the external cocoon, might be realized 
if the cocoon energy close to the equatorial plane is much smaller or if the SN energy is much larger in the equatorial plane. Then, the SN shock front could potentially decelerate but break-out of the cocoon before reaching its local velocity and mixing within the cocoon (figure \ref{fig1}).

\section{Thermal and non-thermal emission from the system}

\begin{figure}
\centering
 \includegraphics[width=0.5\textwidth]{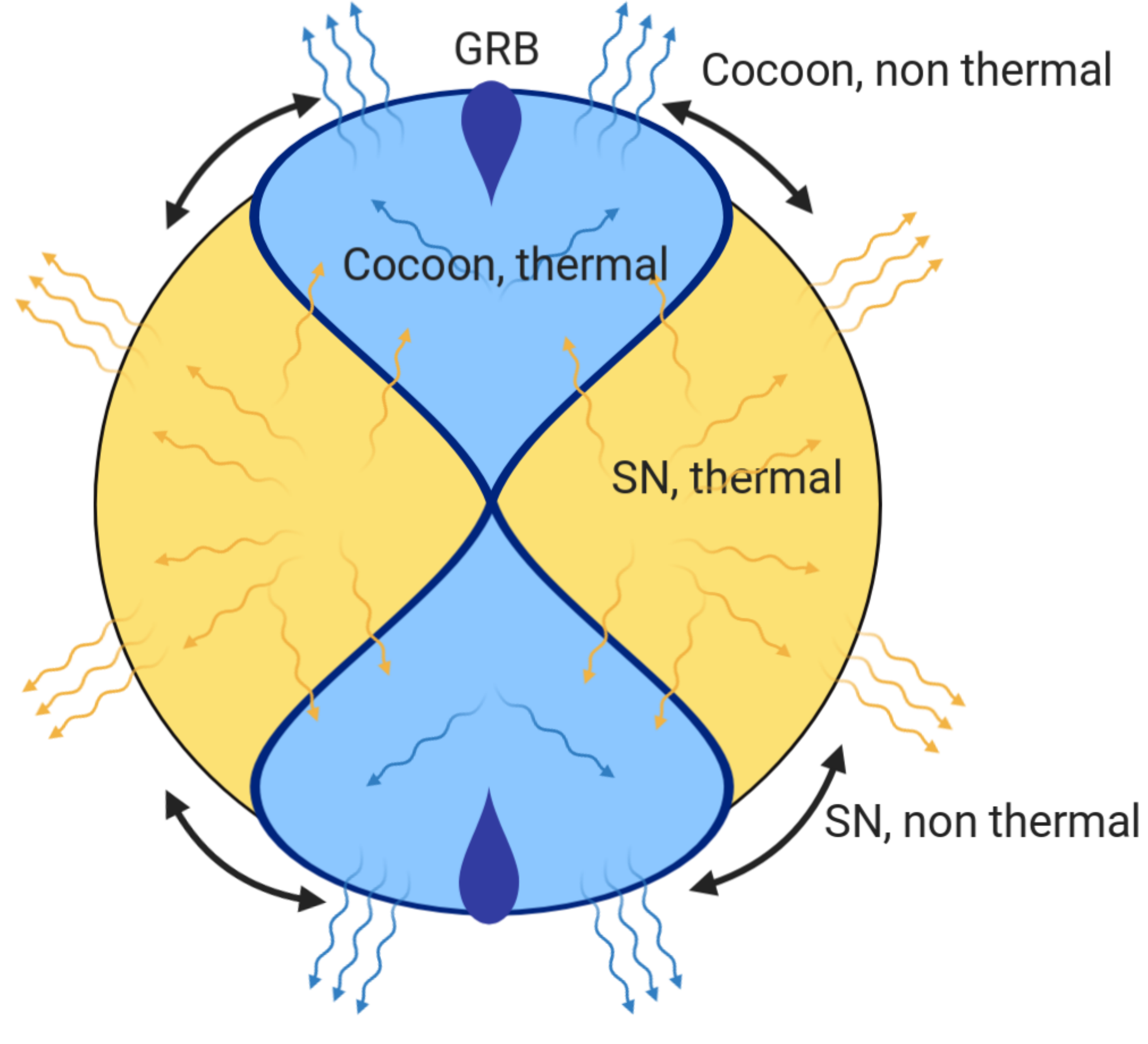}
 \caption{Schematic representation of the emission from the GRB, cocoon and SN ejecta. Both cocoon and SN emit thermal and non-thermal radiation. In case a) (see figure \ref{fig1}), the relativistic jet dissipates its energy inside the epanding SN ejecta. In case b) (see figure \ref{fig1}), the SN shock front expands inside the GRB cocoon. As the jet velocity is much larger than the SN shock front velocity, the SN shock front breaks out eventually only close to the the equatorial plane. Then, thermal photons produced by the SN cross the GRB cocoon when this is seen  on-axis, while they might arrive directly to the observer when it is nearly perpendicular with respect with the jet direction of propagation.}
 \label{fig5}
\end{figure}

Figure \ref{fig5} shows a schematic representation of the emission coming from the the SN ejecta, the relativistic jet and its cocoon. As discussed before, the angular extension of the cocoon (represented in figure \ref{fig5} by curved black arrows) depends on the SN energy, on the progenitor star, on the properties of the relativistic jet and on the delay between the jet and SN launching. The two extreme cases are:  1) the GRB cocoon (associated to a successful or failed GRB) completely engulfing the SN ejecta; 2) the SN shock front covering most of the solid angle. Anyway, we notice that all intermediate cases are in principle possible. 

A variety of observational signatures can be associated to the interaction of SNe and GRB cocoon, covering all electromagnetic spectrum (i.e., from radio to $\gamma$-ray emission).
Non-thermal synchrotron emission comes from the GRB shock front (in case of a successful GRB), the cocoon and the SN shocks expanding through the wind of the progenitor star, and from the SN shock eventually moving through the cocoon material (see figures \ref{fig1} and \ref{fig5}).    
Thermal emission from the cocoon is short lasting, peaking at $\sim$ hundred of seconds after the cocoon break-out, in the X-ray band, and lasting much longer ($\sim$ days) in UV and optical. Depending on the solid angle occupied by the cocoon, the SN thermal emission will by scattered by the cocoon low-density material. Also, the SN will be energized by hot photons coming from the cocoon. In this section, we give a qualitative description of these components. A more detailed calculation is left for a future work.

\subsection{Non-thermal emission}

We assume that a fraction $\chi_e$ of the shocked electrons is accelerated at relativistic speeds by the shock front, creating a population $n_e(\gamma)\propto \gamma^{-p}$ of accelerated electrons. We also assume that a fraction $\epsilon_e$ and $\epsilon_B$ of the post-shock thermal energy $e_{\rm th}$ ends up in the energy of the accelerated electrons and in the post-shock magnetic field energy density, i.e., $\epsilon_e = e_{\rm acc}/e_{\rm th}$ and $\epsilon_B = B^2/8\pi e_{\rm th}$.
The synchrotron emission depends on these three parameters and on the post-shock thermal energy $e_{\rm th} \propto \rho_{\rm amb} (v_{\rm sh}-v_{\rm amb})^2$; where $v_{\rm sh}$ the shock velocity and $\rho_{\rm amb}$ \& $v_{\rm amb}$ the density and velocity of the environment. If the environment is shaped by the wind of the progenitor Wolf-Rayet star, it is expected that $v_{\rm amb} \sim 10^8 \;{\rm cm \; s^{-1}} \ll v_{\rm sh}$. On the other hand $v_{\rm amb} \lesssim v_{\rm sh}$ if the SN shock front is propagating through the cocoon.

As discussed in section 2.2, the SN shock front velocity depends on the structure of the progenitor star, with $v_{\rm sh} \propto t^{-0.3}-t^{-0.1}$. Nevertheless, as shown by the numerical simulations (see figure \ref{fig3}), the presence of a cocoon can affect the propagation of the SN shock front inside the star, modifying the ejecta structure and its deceleration in the external environment, specially at the interface with the cocoon. 

For a decelerating spherical blast wave (i.e. the GRB jet at least during the late afterglow emission), $v_{\rm sh}$ is completely determined by the explosion energy and the stratification of the environment, as $v_{\rm sh}\Gamma_{\rm sh} \propto (E/\rho R^3)^{1/2}$. This is not the case for the expanding cocoon. Similar to the SN case, the unshocked cocoon material crosses the reverse shock continuously energizing it. In addition, the cocoon is strongly structured both in energy and density along the polar direction. For instance, numerical simulations presented by \citet{decolle18a} showed that the cocoon velocity evolves as $v_{\rm sh} \approx t^{-0.05}$ at $45^\circ$ and $v_{\rm sh} \approx t^{-0.1}$ at 90$^\circ$.

\begin{figure}
\centering
 \includegraphics[width=0.45\textwidth]{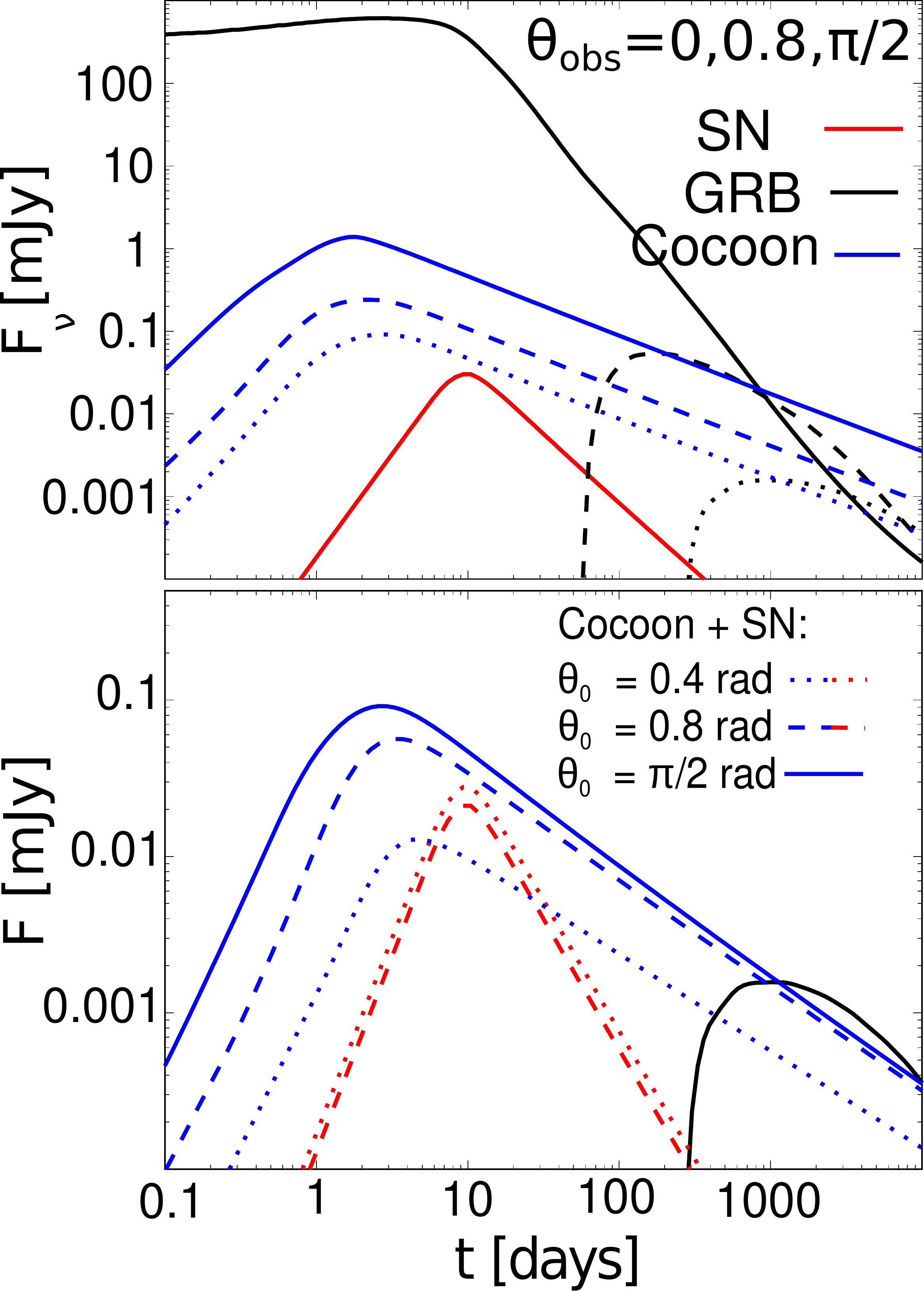}
 \caption{Light curves at 8.46 GHz produced by the GRB (black lines), cocoon (blue) and SN (red). \emph{Upper panel:} the curves correspond to a LGRB located at a distance of 200 Mpc, seen at $\theta_{\rm obs} = 0, 0.8, \pi/2$ (full, dashed, dotted line). The SN emission is taken as spherically symmetric, so that all observing angles see the same flux. \emph{Bottom panel:} curves corresponding to a LGRB located at 200 Mpc, seen by an observer located at $\theta_{\rm obs} = \pi/2$. $\theta_0$ correspond to the angle separating the cocoon from the SN, i.e. $\theta_0=0.4$ rad indicates that the cocoon and SN occupy a region $\theta\leq 0.4$ rad and $\theta\geq 0.4$ respectively. The microphysical parameters used to compute each curve are reported in the main text.}
 \label{fig6}
\end{figure}

Figure \ref{fig6} shows an example of radio light curves produced by a LGRB, its cocoon and the SN as seen by observers located at different angles $\theta_{\rm obs}$ with respect to the direction of propagation of the relativistic jet. The light curves have been computed by considering a decelerating top-hat jet \citep{decolle12b}, the cocoon associated to it  \citep{decolle18a} and a typical SN light curve (following the analytical prescription of \citealt{chevalier98}).

We consider a relativistic jet with an isotropic energy $E_{\rm iso} = 10^{53}$ ergs  moving into a windy medium with mass-loss rate $\dot{M}_w =2\times 10^{-6}$ M$_\odot$ yr$^{-1}$ and a wind velocity $v_w=10^8$ cm s$^{-1}$ \citep{decolle12b}. The jet is initialized as a wedge with an opening angle $\theta_j=0.2$ rad with a shock Lorentz factor of $20\sqrt{2}$. The post-shock density, pressure and velocity are given by the \citep{blandford76} self-similar solution.
The cocoon velocity structure is taken from the numerical simulations presented by \citet{decolle18a}. In these simulations, we launched a jet with a total energy of $2\times 10^{51}$ ergs, an opening angle $\theta_j=0.1$ and a duration $t_j=10$~s. The progenitors star is the same used in the simulations presented in this paper (the E25 model of \citealt{heger00}). The computational box extends up to $10^{16}$ cm along the r and z-axis, and is resolved by  $60\times 60$ cells at the coarsest grid and 26 levels of refinement. The microphysical parameters are $\epsilon_e=10^{-1}$ and $\epsilon_B=10^{-3}$, $p=2.2$ for both the GRB and the cocoon. These values are typical of GRBs \citep{panaitescu02,santana14}.
The SN shock front is assumed to move with a velocity $v_{\rm sh} = 10^{10} \; (t/{10 \; \rm days})^{-0.2}$ cm s$^{-1}$, $\epsilon_e=0.3$, $\epsilon_B= 0.3$ (i.e., equipartition) and $p=3$ through the same windy medium. These are typical values of the microphysical parameters usually considered when interpreting observations of radio SNe \citep[see, e.g.][]{chevalier17}.
Self-absorption is not included in the calculation of the GRB light curve (it can be important at times $\lesssim 1$ day) but is included in the calculation of the cocoon and SN light curves.

Non thermal emission from the GRB has been extensively studied. It is strongly beamed along the direction of propagation of the relativistic jet. For off-axis observers, the GRB radiation will enter in the field of view when $\Gamma_j \lesssim 1/\theta_j$; where $\Gamma_j$ and $\theta_j$ are the Lorentz factor and opening angle of the highly relativistic moving material \citep[e.g.,][]{granot18}.

While the GRB dynamics during the deceleration phase is in general well described by the \citet{blandford76} self-similar solution (except for the lateral expansion happening when $\Gamma_j\lesssim 1/\theta_j$), the dynamics of the cocoon is more complex, as it is regulated by the energy stratification of the ejecta. Along each polar direction, each shell of the cocoon decelerates at a different deceleration radius, given by \citep{hotokezaka15} $M(R) (v\Gamma) = E(\geq \beta\Gamma)$, where $M(R)$ is the mass included up to the distance $R$ from the central engine, and $v(\Gamma)$ are the velocity/Lorentz factor of the cocoon  (but notice that lateral expansion is more important in GRB cocoons as they move at mildly/sub relativistic speed at all times - as in the GRB case, its effect can be captured precisely only by numerical simulations). 

The cocoon energy is determined by the jet break-out time (i.e., $E_c \sim L_j t_b$), while the  jet total energy is given by $E \sim L_j (t_{f}-t_b)$, being $t_f$ the total injection time of plasma from the central engine, $t_b$ the breakout time and $L_j$ the jet luminosity (assumed constant). 
It is expected that high luminosity LGRB emission dominates when seen on-axis with respect to the cocoon emission, while the bolometric cocoon emission can be larger than the GRB emission at late times (see the upper panel of Figure \ref{fig6}) if the total energy of the cocoon is larger than the GRB energy, potentially producing a flattening of the light curve $\sim$ months - years after the explosions. 
Then, the cocoon emission is expected to be more prominent in long GRBs lasting for a shorter time or in progenitors with large radius, in which the jet emerges from the surface of the progenitor star with only a small fraction of its initial energy.

A flattening of the radio light curve has been observed in several GRBs and interpreted as evidence of the deceleration of the jet to non-relativistic speeds \citep{frail04, chen20}. Our results suggest that an alternate explanation might be the onset of the cocoon emission at late times.

On the other hand, the cocoon dominates the emission on timescales of 10-1000 days (depending on the observing angle and on the density of the ambient medium) for off-axis GRBs (consistent with observations of GRB 170817A), specially if the jet is seen nearly perpendicular to its axis. Also in this case, the light curve can present one or two peaks depending on the GRB and cocoon characteristics and on the observing angle.

We also notice that a different choice of the microphysical parameters, e.g., with $\epsilon_e$ and $\epsilon_B$ increasing as the shock velocity drops (consistently with the gradient of values observed when considering relativistic GRBs and non-relativistic SNE), could increase the importance of the cocoon possibly producing a double peak structure (see figure 6 of \citealt{decolle18a}).
The SN emission is always negligible in this case in radio frequencies.

Non thermal emission from the GRB cocoon is mildly beamed at early stages when the cocoon moves with Lorentz factors of $\sim$ a few, and isotropic at later stages as the cocoon velocity drops with time  \citep[see, e.g.,][]{decolle18a}.
In figure \ref{fig6}, bottom panels, we consider the different cases shown in Figure \ref{fig5}. In our calculations, the SN radio emission peaks at $\sim 10$ days, while the cocoon emission peaks at $\sim 1-10$ days (once the region closer to the jet core slows down and becomes visible). The cocoon and SN can cover different solid angle fractions. Then, the SN peak is observable if the SN shock front covers most of the solid angle (i.e., if the SN breaks out from the progenitor star before or a few seconds after the GRB cocoon). Although the importance of each component depends on the energy distribution of each component and on the density stratification of the environment, our results show that, at least for some combinations of the physical parameters, one, two or three peaks could be observed when a GRB is observed off-axis. Future observations of synchrotron radiation from off-axis GRBs will then provide information on the SN and jet ejection processes.

An additional source of non-thermal synchroton emission can be produced by the propagation of the SN shock front through the cocoon. The cocoon density can be several orders of magnitude larger than the density of the wind of the progenitor star at the same radius. The radio emission strongly depends on the velocity difference between the SN shock front and the cocoon. 
During the free-expansion phase, the cocoon and SN velocities are given as $v_c(r,t) = r/t$ and $v_{\rm SN}(r,t) = r/(t-\Delta t)$, while the cocoon and SN shock fronts are given as $R_c = V_c t$ and $R_{\rm SN} = V_{\rm SN} (t-\Delta t)$, being $\Delta t$ the time delay between the cocoon and the SN break out from the progenitor star. Thus, the cocoon velocity at the position of the SN shock front is $v_c(R_{\rm SN},t) = R_{\rm SN}/t$ and the velocity difference is
\begin{equation}
  V_{\rm SN} - v_c = V_{\rm SN} \frac{\Delta t}{t} \;,
\end{equation}
for $t > \Delta t$. At early times, $t\gtrsim \Delta t$, and the velocity difference is $\sim V_{\rm SN}$, i.e., the SN collides with the slowing moving cocoon component. As $t$ increases, the SN shock front expands through cocoon material which moves with increasing velocity. Then, the thermal energy in the SN post-shock region (i.e., the energy available to produce radiation) quickly drops with time ($\propto (V_{\rm SN} - v_c)^2\propto t^{-2}$).

The cocoon is radiation-pressure dominated, with a sound speed (computed from numerical simulations) $c_s \sim \sqrt{p/\rho c^2} \sim 0.01\; c$ in the equatorial plane, and $\sim c/\sqrt{3}$ at small polar angles. As the SN shock front expands mainly along the equatorial plane, the Mach number is, after the SN stellar break-out, $\gtrsim 5$, and particles can be accelerated efficiently. On the other hand,
the SN radio emission is typically strongly self-absorbed (see the SN emission in the upper panel of Figure \ref{fig6}), and the SN shell is optically thick until $\sim 100$ days.
In the self-absorbed part of the spectrum, the flux is 
\begin{equation} 
  F_\nu \propto R^2 B^{-1/2} \nu^{5/2}\;.
\end{equation}
Assuming a constant $\epsilon_B$ (i.e., a magnetic energy density proportional to the thermal energy density), we get $B =  8\pi (\epsilon_B e)^{1/2}$, 
$e\propto \rho_c (V_{\rm SN}-v_c)^2 \propto \rho_c V^2_{\rm SN}\Delta t/t^2$, being $\rho_c$ the cocoon density, and 
\begin{equation}
    F_\nu \propto R^2 \left(\rho_w V^2_{\rm SN}\right)^{-1/4} \nu^{5/2}
    \left(\frac{\rho_c}{\rho_w} \frac{\Delta t}{t^2}\right)^{-1/4}\;,
\end{equation}
where the last term in parenthesis represents the correction to the flux emitted by a SN shock front moving through the wind of the progenitor WR (represented by the other terms in the equation).
As the density of the cocoon is several orders of magnitude larger than the density of the WR wind at the same distance, the flux is strongly reduced until $t\gg \Delta t$. At late times, the emission from the SN interacting with the cocoon is larger than the corresponding emission from a SN moving through the wind of the progenitor star only if the shock remains strong. For instance, in the simulation shown in figure \ref{fig4}, the SN shock front becomes subsonic after $\sim 50$ s, thus suppressing any particle acceleration. We conclude that the non-thermal emission from the SN interacting with the cocoon is unlikely to be observable when the SN shock front breaks out from the star in the dense jet cocoon material, although detailed calculations are needed to confirm this result.

\subsection{Thermal emission}

In this subsection, we describe qualitatively the thermal emission from the GRB cocoon, the expected changes in the SN light curves due to the presence of the GRB and its associated cocoon, and the modification of the geometry of the SN ejecta which, in principle, can be inferred from polarization observations. 

The cocoon thermal emission has been studied analytically \citep{nakar17} and by numerical simulations \citep[]{suzuki13, decolle18b}.
Numerical studies have shown that the cocoon emits a short-lived, nearly thermal spectrum peaking in the soft X-rays (corresponding to a temperature of $\sim 1$ keV, see figure \ref{fig4}). Its light curve lasts for $\sim$10 minutes in the observer frame. The photon diffusion timescale across the cocoon is \citep{arnett79} $t_d \sim (M_c k/v_c c)^{1/2}\sim 10^4$ s, being $M_c\sim 10^{-2} M_\odot$ the cocoon mass, $k$ the opacity per unit mass, and $v_c$ the average cocoon velocity over time. As the cocoon is initially moving at mildly to sub-relativistic speeds, the observer time is related to the lab frame time as $t_{\rm obs} = t_d - R/c \sim t_d/2\Gamma_c^2 \sim 10^2-10^3$ s, which is then the expected timescale for the cocoon X-ray thermal emission. As the cocoon expands, it slows down. The X-ray emission will be followed by emission in UV and optical on a timescale of $\sim$ days (Nakar \& Piran 2017), as inner, slower and colder regions are exposed during the cocoon expansion.

The X-ray light curve and spectrum can be  observed only if associated to intrinsically low-luminous GRBs or to GRBs seen off-axis \citep{decolle18b}. SN 2017iuk/GRB 171205A show strong evidence of this emission (\citealt{izzo19}), with a thermal X-ray flux consistent within a factor of $\sim$ a few with the numerical models.
This general picture is modified by the presence of the SN. As the cocoon is mildly relativistic during the thermal emission phase, the emission is likely to be suppressed along the equatorial plane if the SN shock front occupies a large fraction of the solid angle (see figure \ref{fig5}). Observations of a sample of off-axis GRBs and of their thermal X-ray emission will then shed light on the structure of the system.

In general, in core collapse supernovae the shock front breaking out from the progenitor star consists of the fluid shock moving at $\sim 0.1-0.3$ c and two precursors, namely the ionization front with a speed related to the energy flux and the light front moving with the speed of light \citep{mair92}. For the cores of massive star, color temperatures at shock breakout reach several million degrees resulting in a brief X-ray flash followed by softening and UV radiation (Katz et al. 2010). 

While typical Wolf-Rayet winds (with $\dot{M} \sim 10^{-5}-10^{-6}$ M$_\odot$) yr$^{-1}$ have optical depths $\lesssim 1-10$ at the stellar surface, and thus optically thin to X-ray and UV emission at the shock break-out radius, the cocoon remains optically thick up to large distances. Assuming a uniform density distribution in the cocoon, we get $\tau = \sigma_T \rho R/m_h = 3 \sigma_T M_c/(4\pi m_H R^2)$. Thus, $\tau<1$, for the cocoon, is obtained at radius $R\gtrsim 10^{14}$ cm. This should be taken as a very rough estimate as the cocoon is strongly stratified in density along the polar direction (see figure \ref{fig3}, central panel), the SN shock breakout is expected to happen at smaller radii along the equatorial plane then along the jet propagation direction. SN shock breakout happening at a larger radius will be brighter (although at a lower energy) and last much longer\footnote{The shock breakout has also a non-thermal component associated with inverse-Compton scatterings of photons, and possibly some contribution from synchrotron radiation.}.

Velocity differences between the SN shock front and the GRB cocoon produce strong shocks and free-free emission. During the optically thick cocoon expansion phase, this radiation will represent an extra heating source in the SN ejecta. Once the cocoon becomes optically thin (at $R\gtrsim 10^{12}-10^{14}$ cm depending on the cocoon mass), these X-rays may backshine on the photoshere, leading to a boost of the optical luminosity and may lead to a slowly rising LC prior to maximum light \citep[e.g.,][]{dragulin15,hsiao20}.
The light curve associated to bipolar SN explosions can then decline quickly for an observer located along the equatorial plane \citep{kaplan20, soker21}.

Finally, we notice that the geometry of the SN ejecta is strongly modified by its interaction with the GRB cocoon. 
When the SN breaks out in the GRB cocoon, pressure and density gradients make the break out very asymmetric (see Figure \ref{fig1}). Polarization measurements of $\approx 2-3 \% $\footnote{Note that $P\propto P(\max) \sin^2\theta$ being $\theta $ the observer angle.} have consistently shown that the explosion results into strongly asymmetric cores with a dominant axial-symmetric component with axis ratios of about 2 \citep{hoflich91,wang96,hoeflich96,wang01,leonard02,maund07,tanaka17}.
A similar evidence of asymmetry has been obtained by emission by forbidden lines during the nebular phase of SNe \citep{taubenberger09}. Our results suggest as an alternative to an asymmetric explosion that, in massive stars, the interaction between GRB environment and the SNe envelope may produce the asymmetries observed. 

\section{Discussion and conclusions}

In this paper, we described the rich landscape resulting from the interaction of the SN, the GRB and its cocoon. As shown in figure \ref{fig1}, the outcome of the system depends on the SN and cocoon breakout time from the stellar progenitor, and on the timescale for the cocoon to engulf completely the progenitor star. In the case of weak jets, or jets injected with a large time delay with respect to the SN, the SN expands into the progenitor star and the surrounding medium, with the jet being drowned inside the SN ejecta, or breaking out at a much later stage (see figure 1, panel a). In this case, most of the cocoon energy is deposited inside the SN ejecta. Mixing between the jet cocoon plasma and the surrounding medium then increases the average velocity of the SN ejecta. Also, in the case, even when the SN explosion is spherically symmetric, the SN ejecta will become asymmetric due to this interaction. We expect that the thermal emission will be characterized by broad absorption lines, while non-thermal synchrotron emission from the SN shock front will be observed at radio frequencies. These characteristics are common to broad-line, type Ic SNe not associated with GRBs, which can be drowned into the SN ejecta or can be pointed in a direction away from the observer, thus remaining undetected.

A very different outcome is expected when the relativistic jet and its surrounding cocoon are the first to arrive at the progenitor star surface. This is expected to happen when the jet and the SN are launched approximately at the same time, and the jet luminosity per solid angle is at least of order of the SN kinetic luminosity (also per solid angle). In this case, the GRB-jet cocoon will expand directly into the progenitor wind. The SN shock front might break out from the stellar progenitor into the wind or into the GRB cocoon. In the latter case, the SN shock front will possibly break out from the GRB cocoon at a later time, being the final outcome of the system in this case mainly regulated by the cocoon structure once the SN breaks-out from the progenitor star.
The cocoon structure depends on the stellar structure, with more compact stars producing more massive cocoons, and on the jet characteristics, e.g., opening angle, luminosity, etc. which regulates the amount of mass and energy deposited into the cocoon.
Magnetically dominated jets, in particular, are expected to present cocoon with lower mass and energy which will be more easily swapped up by the SN ejecta.

 The cocoon shock front is expanding initially at relativistic speeds, with lorentz factors of $\sim$ 5-10, then decelerating to sub-relativistic speeds. Synchrotron emission associated with mildly relativistic material has been observed in relativistic SNe, and it is naturally expected in all systems in which the cocoon is propagating directly through the progenitor wind (see figure \ref{fig6}). We showed that the cocoon emission can also explain the late time flattening observed in several cosmological GRBs, as the cocoon emission will be larger than the GRB emission at late times, specially for short duration long GRBs (this prediction can be easily checked observationally). The SN shock front moving inside the cocoon is also expected to produce non-thermal emission. While for the initial conditions employed in our simulations we showed that this emission is short-lived and probably undetectable, a more complete exploration of the parameter space is needed before drawing firm conclusions. In particular, in the case of a less dense cocoon (with respect to that obtained in the simulations shown in this paper), the SN shock front is expected to travel much longer distances before mixing with the cocoon material, or possibly breaking out from it.

Thermal emission from the SN is affected in a few different ways by the presence of the cocoon and GRB: 1) the presence of highly pressured cocoon makes the SN propagation asymmetric (see figure \ref{fig3}), which will have an affect on the SN polarization. 2) The GRB cocoon can be denser than the progenitor wind at the same radius. Thus, the interaction between the SN ejecta and the cocoon produces free-free emission, which represents an extra energy source in optically thick regions, and can be detected in X-rays once the cocoon becomes optically thin. 3) The cocoon is optically thick when the SN shock front breaks out from the progenitor star surface. As a result, the radiation from the shock breakout will be delayed, and the breakout flash we see is released when SN shock front has moved to a much larger distance.

Finally, we stress that the model presented in this paper is a natural consequence of the ``standard'' LGRB model, in which a relativistic jet is associated with the collapse of a massive star and to the SN production. Although the interaction between SN and GRB cocoon has been studied in the past analytically and numerically, this has been done by considering only their late interaction, i.e. weeks to months after the explosion \citep{ramirezruiz10, margalit20}. In this paper, we showed that it is the early evolution of the system (i.e., while the jet and SN-shock are moving inside the star or have just broken out from it) that mainly determines the fate of the system. Future modeling of the rich landscape described in this paper, and detailed comparison with present and future observations, will test the different scenarios and ultimately provide information regarding the mechanism responsible for the jet and SN creation.

\section*{Data Availability Statement}

The data underlying this article will be shared on reasonable request to the corresponding author.

%
\section*{Acknowledgements}

We thank Luca Izzo for usefull discussions, and the referee, Ehud Nakar, for numerous comments and suggestions that substantially improved the manuscript.
We acknowledge the computing time granted by DGTIC UNAM on the supercomputer Miztli (project LANCAD-UNAM-DGTIC-281). FDC acknowledges support from the UNAM-PAPIIT grant AG100820. PK acknowledges support of the NSF grant AST-2009619. 
Figures \ref{fig1} and \ref{fig5} were created on www.biorender.com.

%
%

\bsp	
\label{lastpage}
\end{document}